\documentclass[journal=nalefd,manuscript=letter]{achemso}
\usepackage{amsmath,amssymb}
\usepackage{graphicx,color,braket,amsmath,dcolumn,bm}

\title{Terahertz photoconductivity in bilayer graphene transistors: evidence for tunneling at gate-induced junctions}
\author{Dmitry A. Mylnikov}
\email{mylnikov.da@yandex.ru}
\affiliation{Center for Photonics and 2D Materials, Moscow Institute of Physics and Technology, Dolgoprudny, 141700, Russia}

\author{Elena I. Titova}
\affiliation{Center for Photonics and 2D Materials, Moscow Institute of Physics and Technology, Dolgoprudny, 141700, Russia}
\alsoaffiliation{Programmable Functional Materials Lab,
Brain and Consciousness Research Center, Moscow, 121205, Russia}

\author{Mikhail A. Kashchenko}
\affiliation{Programmable Functional Materials Lab,
Brain and Consciousness Research Center, Moscow, 121205, Russia}
\alsoaffiliation{Center for Photonics and 2D Materials, Moscow Institute of Physics and Technology, Dolgoprudny, 141700, Russia}

\author{Ilya V. Safonov}
\affiliation{Center for Photonics and 2D Materials, Moscow Institute of Physics and Technology, Dolgoprudny, 141700, Russia}
\alsoaffiliation{Programmable Functional Materials Lab,
Brain and Consciousness Research Center, Moscow, 121205, Russia}

\author{Sergey S. Zhukov}
\affiliation{Center for Photonics and 2D Materials, Moscow Institute of Physics and Technology, Dolgoprudny, 141700, Russia}

\author{Valentin A. Semkin}
\affiliation{Center for Photonics and 2D Materials, Moscow Institute of Physics and Technology, Dolgoprudny, 141700, Russia}

\author{Kostya S. Novoselov}
\affiliation{Institute for Functional Intelligent
Materials, National University of Singapore, Singapore, 117575, Singapore}
\alsoaffiliation{Programmable Functional Materials Lab,
Brain and Consciousness Research Center, Moscow, 121205, Russia}

\author{Denis A. Bandurin}
\affiliation{Department of Materials Science and Engineering, National University of Singapore, 117575, Singapore}

\author{Dmitry A. Svintsov}
\affiliation{Center for Photonics and 2D Materials, Moscow Institute of Physics and Technology, Dolgoprudny, 141700, Russia}

\begin{document}

\begin{abstract}
Photoconductivity of novel materials is the key property of interest for design of photodetectors, optical modulators, and switches. Despite the photoconductivity of most novel 2d materials has been studied both theoretically and experimentally, the same is not true for 2d $p$-$n$ junctions that are necessary blocks of most electronic devices. Here, we study the sub-terahertz photocoductivity of gapped bilayer graphene with electrically-induced $p$-$n$ junctions. We find a strong positive contribution from junctions to resistance, temperature resistance coefficient and photo-resistivity at cryogenic temperatures $T \sim 20$ K. The contribution to these quantities from junctions exceeds strongly the bulk values at uniform channel doping even at small band gaps $\sim 10$ meV. We further show that positive junction photoresistance is a hallmark of interband tunneling, and not of intra-band thermionic conduction. Our results point to the possibility of creating various interband tunneling devices based on bilayer graphene, including steep-switching transistors and selective sensors.

\end{abstract}

\maketitle

Understanding the interplay between optical and electrical properties of novel materials is important for design of various optoelectronics devices, including optical modulators~\cite{Moduators_conducting_oxides,Modulators_review} and switches~\cite{EO_switch_graphene_oxide,EO_switch_liquid_crystal}, photo-transistors~\cite{Phototransistor_2d_double_heterojunction,Phototransistor_negative_cap,Phototransistor_organic} and photo-detectors~\cite{Photodetectors_2d_review,Photodetectors_2d_review_2}. The phenomenon of photocondcutivty (PC), the change in conduction upon illumination, enables to extract the information about fundamental materials' properties, such as band gaps and spectra~\cite{Teppe_PC_ZeroFieldSplitting,Teppe_PC_TopologicalTransition}, and collective excitations~\cite{CR_overtones,Transport_detection_of_collective_excitations}. On the other hand, PC has enabled the practical implementation of detectors in most electromagnetic ranges~\cite{HEB_superconductors,Bolometers_review,market}. The search for photoconductive materials is nevertheless continued, mainly to ensure fast photoresponse and sensitivity in terahertz (THz) frequency range where absorption of most semiconductors is blocked by the band gap~\cite{THz_detectors_review}.

Among numerous novel 2d materials, graphene and its bilayer (BLG) attract a special attention in THz detection applications~\cite{bandurin_resonant_2018,castilla_2019,gayduchenko_2021,Efetov_BLG_calorimeter}. The selective light-induced heating of electronic sub-system in carbon layers ensures fast, potentially sub-picosecond, photoresponse~\cite{tielrooij_2015}. The gate-induced doping and electric tunability of band structure enable the induction of light-sensitive $p$-$n$ junctions without any chemical modifications~\cite{Ryzhii_graphene_pin-detectors}. Though the highly-sensitive light detection by electrically-induced junctions in graphene~\cite{castilla_2019} and its bilayer~\cite{gayduchenko_2021} have been reported, the origins of photoresponse and even transport through such junctions have yet to be understood. The complexity of transport through junctions in BLG stems from comparable values of band gap, Fermi energy, temperature, and disorder amplitude. Depending on sample quality and temperature, the transport through such junctions may be dominated by thermionic emission, variable-range hopping, or tunneling~\cite{Oostinga_bandgap_opening,alymov_2016,Levitov_oscillatory_tunneling,VRH_conduction_BLG_1,VRH_conduction_BLG_2,Shklovsky_VRH_2d}.

\begin{figure*}
\includegraphics[width=\textwidth]{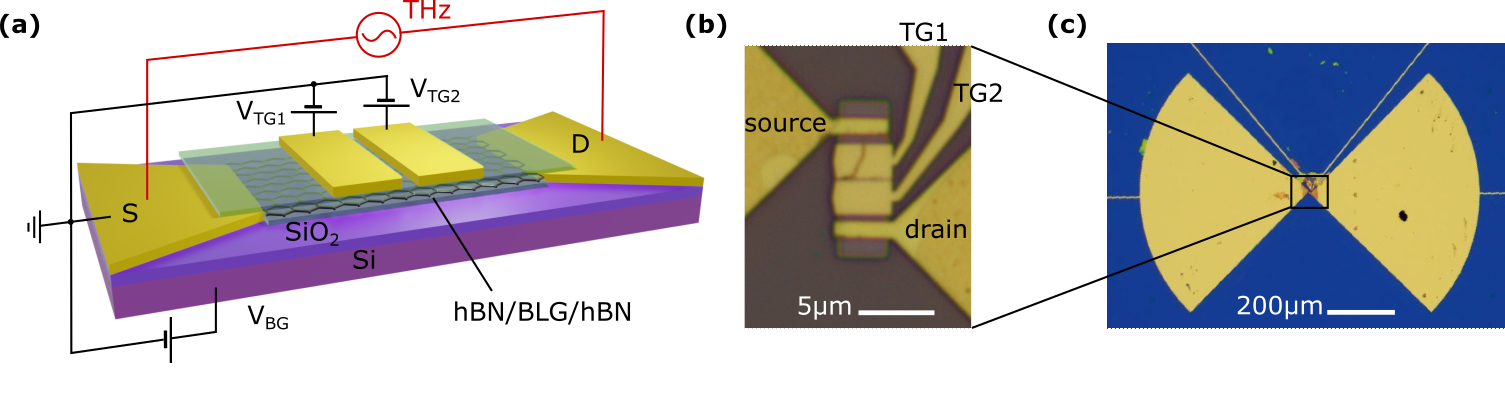}
\caption{(a) Schematic representation of split-gate BLG device. Illuminated antenna connected to source and drain acts as alternating voltage source. Top gate voltages $V_{TG1}$ and $V_{TG2}$ define an in-plane $p$-$n$ junction, while the back gate voltage $V_{BG}$ induces the gap (b,c) Micro-photographs of the fabricated device: channel with split gate (b) and bow-tie antenna (c).}
\label{fig1}
\end{figure*}

Here, we clarify the physics of transport in electrically-induced BLG $p$-$n$ junctions via combined experimental and theoretical studies of terahertz photoconductivity. The structure under study represents a bilayer channel with global back gate and split top gate. The gating architecture enables to switch between gapless and gapped states in the channel, and between uniform and ambipolar doping profiles simultaneously. Previously, the photoconductivity induced by electron heating in graphene bilayer has been studied only in 'junctionless' state, and high bolometric responsivity was demonstrated therein~\cite{yan_2012}. Our study shows that PC signal is strongly enhanced once the $p$-$n$ junctions in the channel are induced. Moreover, the observed negative PC is opposite to that in conventional junctions, where light enhances the conduction. We show theoretically that positive photoresistance induced by electron heating appears if the transport at the junction is dominated by interband tunneling~\cite{alymov_2016,gayduchenko_2021}. Indeed, light-induced heating increases the electron kinetic energy and pushes them out of the 'tunneling window' where conduction and valence bands overlap.

Our devices were fabricated by encapsulating the bilayer graphene (BLG) between two slabs of hexagonal boron nitride (hBN) of thickness 60~nm and 40~nm for top and bottom slabs using a standard dry transfer technique~\cite{kretinin_2014} on top of silicon wafer with 280~nm SiO$_2$. Etching of BLG/hBN slab was used for channel patterning to rectangular shape of width $W=3$~$\mu$m and length $L=6$~$\mu$m. Subsequently, e-beam sputtering was used to deposit top gate and quasi one-dimensional Au/Cr source and drain contacts~\cite{1d_contact}. Finally, plasma etching was used to induce a narrow ($\sim 200$~nm) trench between the gates (Fig.~\ref{fig1}). Further details of sample fabrication detail are given in the Supplementary material, Sec. I. We have studied two devices with similar geometry and obtained identical trends in gate-controlled resistivity and photoconductivity.

It is important that a part of graphene channel between in-plane contact and top gate of $300-400$~nm length is controlled by the back gate only. Therefore, $p$-$n$ junctions in the graphene channel are induced either when voltages of opposite polarity are applied to top gates \#1 and \#2 (mid-channel junction), or when opposite polarities are applied to bottom gate and a pair of top gates (near-contact junction).

To prove the strong effect of gate-induced $p$-$n$ junctions on charge transport, we start with electrical measurements. BLG resistivity was measured with a conventional lock-in technique by passing alternating current of amplitide $I_{\rm bias} = 280$~nA. The device was mounted in an optical cryostat and cooled down to $T=25$ K temperature measured directly on silicon chip. This temperature, if not stated otherwise, is assumed in all further results.

\begin{figure*}
\includegraphics[width=0.8\textwidth]{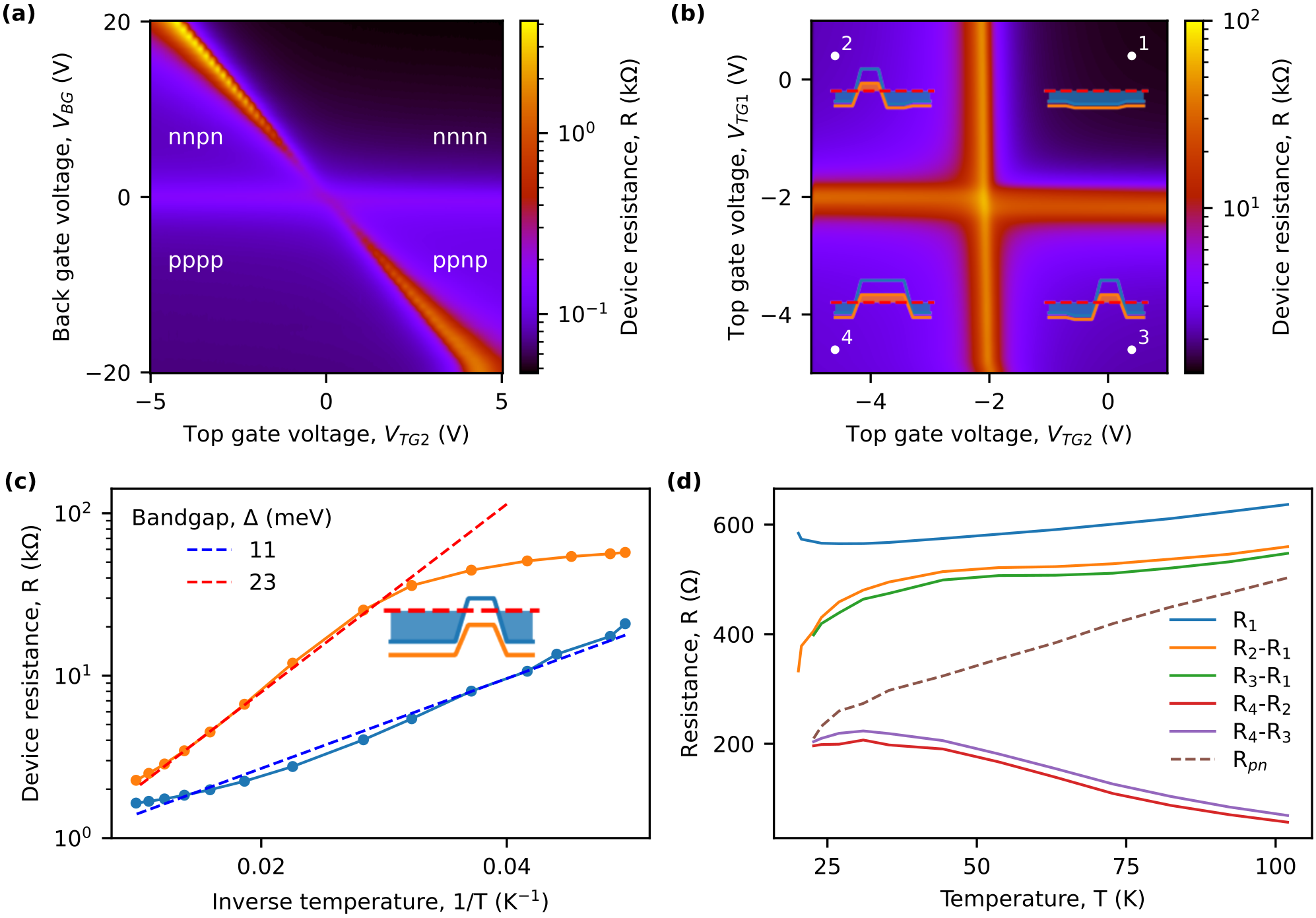}
\caption{(a,b) Resistance maps of the BLG device at $T=25$~K obtained by sweeping (a) back gate voltage and top gate voltage $V_{TG2}$ at fixed $V_{TG1}=0$ (b) two top gate voltages at fixed $V_{BG}=10$~V. Insets in (b) show the band diagrams at biasing conditions marked by points 1-4 (c) Arrhenius plot of the device resistance recorded at CNP of one of the gated regions (while other regions are heavily doped) for two different $V_{BG}$, blue line for $V_{BG}=10$~V ($V_{TG2}=-2.1$~V, $V_{TG1}=0.4$~V), orange line for $V_{BG}=20$~V ($V_{TG2}=-4.2$~V, $V_{TG1}=-1.7$~V). Dashed lines represent Arrhenius fit $R\sim\exp(\Delta/2kT)$. (d) $T$-dependent resistance of BLG channel at different doping: $n$-doped channel with contacts $R_e=R_1$, resistance due to induced $p$-$n$ junctions  $R_2-R_1$ and $R_3-R_1$, additional resistance of hole-doped channel $R_4-R_2$ and $R_4-R_3$, resistance of $p$-$n$ junctions $R_{pn}$.}
\label{fig2}
\end{figure*}

The measured map of two-point resistance vs voltages at the two gates, $V_{\rm BG}$ and $V_{TG2}$, is shown in Fig.~\ref{fig2}. In agreement with previous measurements~\cite{Oostinga_bandgap_opening,Stampfer_AEM_transport_spectroscopy}, the maximum resistance is achieved at the map diagonal. Its slope, $\Delta V_{\rm BG}/\Delta V_{\rm TG} \approx 0.2$, determines the ratio of gates' efficiency, and coincides well with the ratio of the effective thickness of top and bottom dielectrics. Expectantly, the resistance maximum is achieved at largest absolute values of back gate voltages, which is consistent with gap opening in the channel induced by transverse field. A secondary resistance maximum seen at $V_{\rm BG} \approx 0$ can be attributed to the neutrality condition of regions not covered by the top gate.

A prominent feature of the measured resistance maps is the considerable enhancement of the two-point channel resistance $R$ in case when $p$-$n$ junctions are induced. To prove the junction-related origin, we classify the four characteristic regions in Fig.~\ref{fig2}(a) in accordance with polarity of four characteristic sections in the channel. For example, the notation $nnpn$ (top left part of the map) corresponds to $n$-doping of near-contact areas, $n$-doping under the first top gate and $p$-doping under the second one. Only the uniformly doped region have relatively small resistance guaranteed by the absence of the junctions. A similar asymmetry of transfer curves was seen in several~\cite{Bandurin_EdgeCurrentsShunt,Koppens_Science_VSHE} (not all~\cite{Stampfer_AEM_transport_spectroscopy}) previous measurements of BLG, but scarcely attained proper attention~\cite{gayduchenko_2021}. Slightly higher resistance at the $p$-side, $R_{pppp} > R_{nnnn}$ may be attributed to electron doping of BLG by Au/Cr contacts.


Further insights into junction effects on net resistance are obtained by sweeping two top gate voltages at fixed value of $V_{BG}$. An example obtained at $V_{BG} = + 10$~V is shown in Fig.~\ref{fig2}(b). Clearly, only $n$-doping of both gated regions, identical to doping of near-contact areas, promotes low resistance. Inverting the back gate voltage polarity, one obtains the resistance map with minimal resistance in the left bottom corner (see Supplementary material, Sec. IV). It proves that junctions of interest are formed at the boundary between single- and double-gated regions, but not at the metal-graphene contact.




We further perform temperature-dependent transport measurements providing insights into mechanisms of resistivity. The $T$-dependence of resistance at the charge neutrality point (CNP) follows the Arrhenius law, $R = R_0 \exp\{\Delta/2kT\}$, where $\Delta$ is the bandgap, see Fig.~\ref{fig2}(c). The extracted values of $\Delta$ at $V_{\rm BG}=10$~V and $V_{\rm BG}=20$~V are 11~meV and 23~meV, respectively. These values are in fair agreement with theoretical prediction $\Delta = \alpha |C_{\rm TG} V_{\rm TG} - C_{\rm BG} V_{\rm BG}|/\varepsilon_0$, where $C$ is the gate-channel capacitance per unit area, and $\alpha \approx 0.05$~eV$\cdot$nm/V~\cite{Falko_bandgap_theory}, yielding $\Delta = 13$ and 26~meV, respectively. At the smallest temperatures, $T^* \approx 21$~K, the resistance tends to saturation indicating the leakages through edge states and/or hopping via defect levels~\cite{Bandurin_EdgeCurrentsShunt,Oostinga_bandgap_opening}. 



To single out the effect of $p$-$n$ junctions on $T$-dependent resistivity, we further show the $R(T)$-curves for four characteristic configurations with identical absolute values of carrier density in the bulk. These configurations, $nn$, $pn$, $np$, and $pp$, are marked by points 1-4 in Fig.~\ref{fig2}(c). Their temperature-dependent resistances are shown in Fig.~\ref{fig2}(d). The resistance of uniformly-doped $n$-channel $R_{1}$ is order of 600~$\Omega$ and increases with temperature, which is common for phonon-limited transport. Importantly, the {\it added} resistances of non-uniform channel $R_2-R_1$ and $R_3-R_1$ also grow with temperature. We further extract the $T$-dependent resistivity of a single $p$-$n$ junction $R_{pn}$, which are assumed identical in all regions. The analysis accounts for different bulk conductivities of electrons $R_e$ and holes $R_h$ which can emerge due to different effective masses~\cite{Band_asymmetry}. Under these assumptions, $R_2 - R_1 = 2 R_{pn} + R_h - R_e$ and $R_4 - R_2 = R_h - R_e$, which is sufficient for determination of $R_{pn}$. The result of extraction, shown in Fig.~\ref{fig2}(d) with dashed line, again grows with $T$. This lets us conclude that $p$-$n$ junctions in BLG have a positive temperature coefficient of resistance (TCR). This contrasts to conventional junctions in Si, Ge, and III-V semiconductors.

\begin{figure*}
\includegraphics[width=0.9\textwidth]{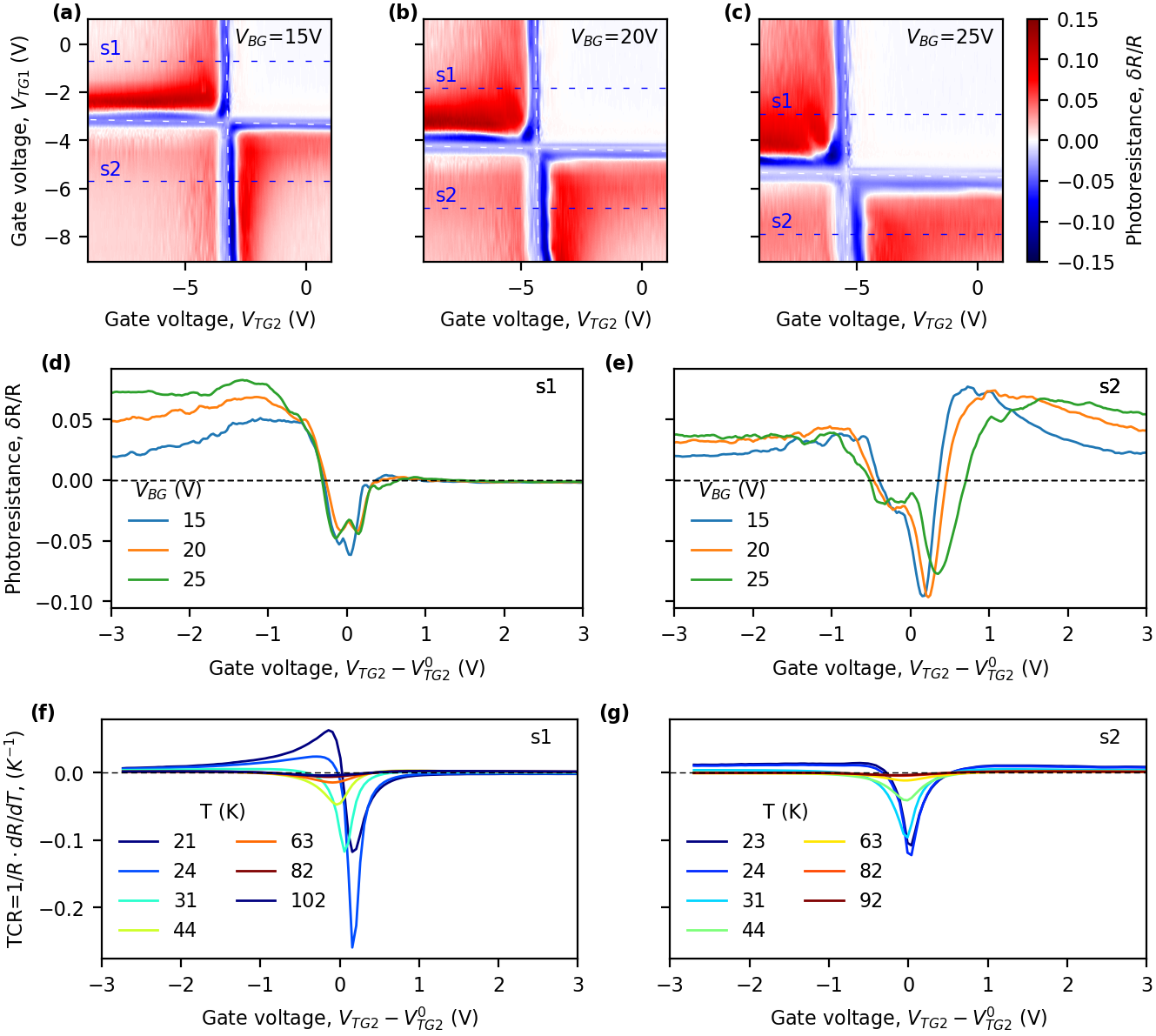}
\caption{(a-c) Maps of BLG device photo-resistance at different back gate voltages indicated in plots. White dashed lines correspond to CNP. (d,e) Slices of photoresistance maps at $V_{TG1}$ 2.5~V above (s1) and below CNP (s2). (f,g) Measured temperature coefficient of resistance at slices (s1) and (s2) recorded at $V_{BG} = 10$~V. In panels (d-g), $V^0_{TG2}$ is the charge neutrality voltage different for each $V_{BG}$}
\label{fig3}
\end{figure*}

Now it is tempting to see whether the measured $T$-dependent resistivity of split-gate BLG device makes any fingerprints on the electro-optical properties. To this end, we measure the device photoconductivity upon illumination with sub-terahertz radiation. During electro-optical measurements, radiation from the sub-THz source (0.13~THz IMPATT diode) was focused to the sample through Si lens glued to the back side of the substrate. Moderate resistivity of substrate (12~$\Omega\cdot$cm, B-doped Si) ensured its transparency. A specially designed wideband THz bow-tie antenna~\cite{mylnikov_2022} was connected to source and drain contacts. The polarization of the light was set along the axis of the bow-tie antenna. The radiation power was chosen so that the device operated in a linear mode, and was 0.5~$\mu$W (Supplementary material, Sec. V). The photoconductivity was measured using a double-modulation technique: the radiation from sub-THz source was modulated at $f_\mathrm{light}=34$~Hz, the while bias current through the sample oscillated at $f_c=111$~Hz. The signal measured at difference frequency $f_\mathrm{light}-f_c$ carries the information about PC.


The measured maps of relative photoresistace (PR) are presented in Figs. \ref{fig3}(a)-(c). Its pattern is highly non-trivial, depending on the carrier density in the channel. First, the photoresistance is negative in the vicinity of CNPs under the first and the second gates (marked by white dashed lines). The width of cross-shaped negative photoresistance area grows with increasing the band gap. Second, the photoresistance changes sign abruptly as we shift either of top gate voltages from its CNP. The only exception of this rule occurs if the channel is uniformly doped. In the latest case, the photoresistance signal is very small, so that even its sign cannot be resolved. Nearly-zero signal for uniformly doped channel persists if we invert the back gate voltage (see Supplementary material, Sec. IV).  




From the first glance, it may be tempting to explain the observed features of THz PR by the variations of bulk temperature resistance coefficient $dR_{\rm bulk}/dT$. Indeed, in the linear regime $\delta R_\mathrm{THz} = (dR/dT) \delta T_\mathrm{THz}$. As radiation always heats the electrons, $\delta T_\mathrm{THz} >0 $, and the sign of photoresistance coincides with that of $dR/dT$. The negative PR near the CNP is then readily explained by thermal generation of electrons and holes as the THz radiation heats the charge carriers. The negativity of PR should persist as far as the Fermi level stays within the gap. This fact is reflected in a wider negative PR cross with increasing the back gate voltage (compare Figs.~\ref{fig3}(a) and (c)). The positive PR away from CNP is typically explained by phonon-limited transport in the 'metallic state', where $dR/dT$ is positive~\cite{Pump-probe_PC_BLG,Pump-probe-PC-BLG2}. However, in our case, this explanation fails. Indeed, this scenario does not explain the asymmetrically  vanishing photoresistance for uniformly doped channel.

We now argue that strong positive PR of non-uniformly doped BLG is a consequence of peculiar $T$-dependent resistance of $p$-$n$ junctions. To this end, we compare the measured values of $\delta R_\mathrm{THz}/R$ and $dR/dT$ along specific slices $s_1$ and $s_2$ of the full map. The data at slices are shown in panels (d)-(g) of Fig.~\ref{fig3} and cover all four combinations of doping ($nn$, $np$, $pn$, and $pp$). One observes that positive PR correlates well with positive $dR/dT$ in the $np$, $pn$, and $pp$ regions away from CNP. The absolute value of $dR/dT$ in the $nn$-region, where junctions are absent, is smaller, and the sign of $dR/dT$ cannot be unambiguously resolved.

\begin{figure}
    \centering
    \includegraphics[width=0.5\textwidth]{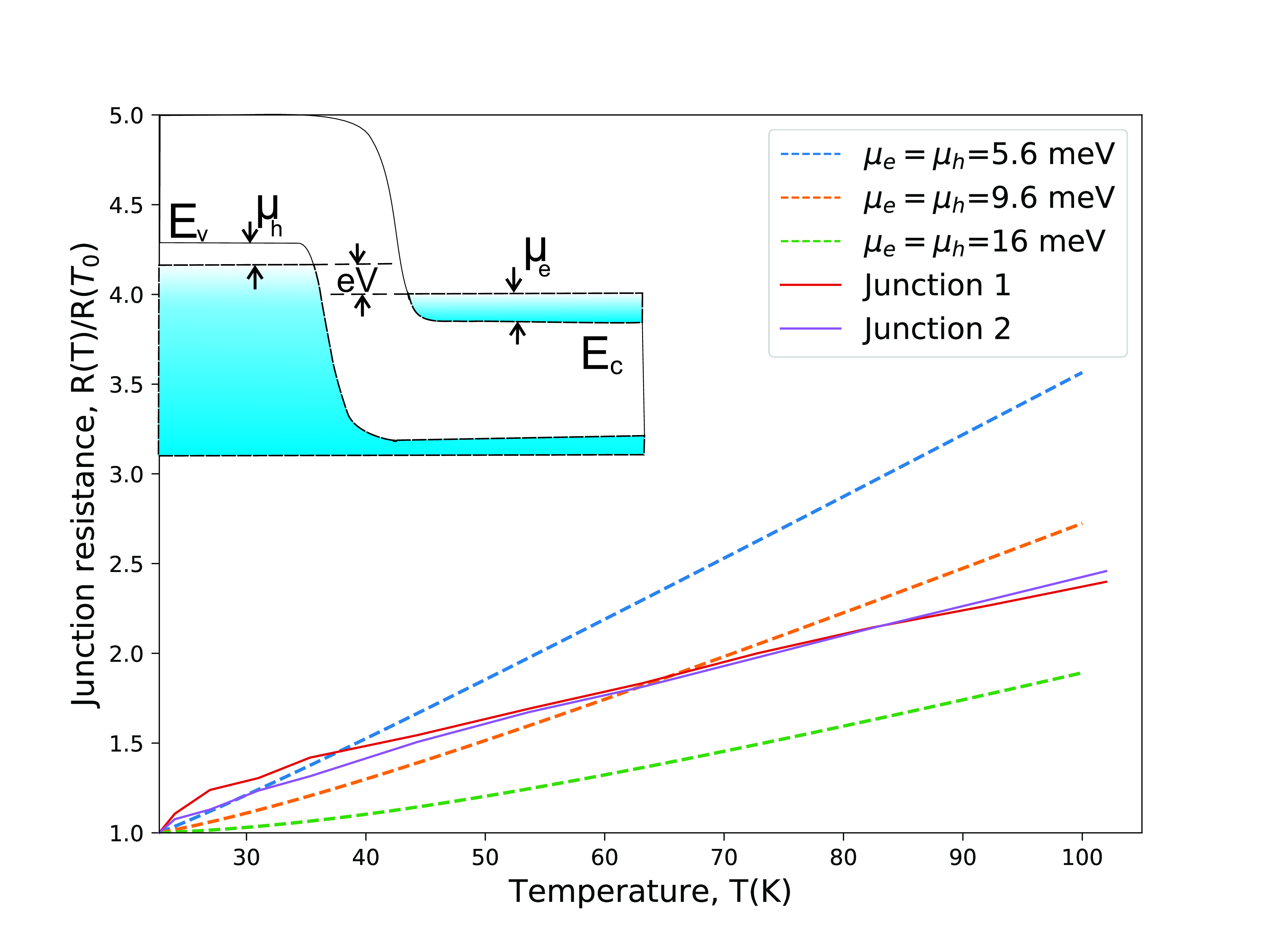}
    \caption{Calculated dependences of tunnel $p$-$n$ junction resistance on temperature (dashed lines) along with experimentally extracted resistances of two junctions (solid lines). The data is normalized to resistance at the lowest temperature of measurements, $T_0 = 22.5$ K. Different colors correspond to different Fermi energies in the $p$- and $n$-regions (marked in the inset)}
    \label{fig4}
\end{figure}

The simultaneous measurements of $\delta R_\mathrm{THz}$ and $dR/dT$ at various doping profiles in the channel hint on {\it junction-induced} asymmetric photoresistance. Still, the positive sign of PR (and of $dR/dT$) of $p$-$n$ junctions in graphene bilayer raises extra questions. Indeed, if the barrier between $p$ and $n$ sides is overcome by thermally activated carriers, the value of $dR/dT$ should be negative, as it occurs for conventional semiconductor junctions. Even if the conduction occurs via variable-range hopping through defect levels~\cite{Oostinga_bandgap_opening} or electron-hole puddles~\cite{Shklovsky_VRH_2d} (specific to narrow-gap 2d systems), this should also result in negative TCR.

We proceed to show that positive TCR of $p$-$n$ BLG junctions can be explained by a simple interband tunneling theory accounting for finite-temperature smearing of the Fermi surface. The relevant band diagram is shown in the inset of Fig.~\ref{fig4}. Intuitively, the electrons capable of tunneling are constrained in energy by the edges of conduction and valence bands, $E_C$ in the $n$-region and $E_V$ in the $p$-region. Raising the temperature increases the average kinetic energy of electron and 'pushes' them away from the tunneling window $E_V - E_C$. Most of thermally activated electrons are now blocked by the gap and do not contribute to transport, and only a small fraction of thermally activated electrons may overcome the barrier.

Direct calculations of tunneling junction conductance based on Esaki theory confirm this idea. Assuming energy-independent tunneling probability $D_0$ and constant densities of states $\rho_c$ and $\rho_v$ in the conduction and valence band, we arrive at a remarkably simple formula for $T$-dependent tunneling conductance:
\begin{equation}
    G_{\rm tun}(T) \propto D_0 \rho_c \rho_v [f(E_C,T) - f(E_V,T)],
\end{equation}
where $f(E,T) = [1+e^{(E - F)/kT}]^{-1}$ is the Fermi function, and $F$ is the Fermi level (details are in Supplementary material, Sec. VI). The result of theoretical calculations shown in Fig.~\ref{fig4} for various Fermi energies (not known exactly) produces a fair agreement with experiment. Further extension of the model including the energy-dependent density of states in BLG may improve the agreement.

Though the observed dependences of photoresistance on doping can be explained by variations of TCR for the overall FET structure, the variations of THz-induced temperature increase $\delta T_\mathrm{THz}$ may also play role. Particularly, the presence of extra junction resistances in non-uniformly doped channel may block the cooling of hot electrons via leakage to the contacts~\cite{Fonf_PRX_Cooling}. Simple estimates show that an average increase in electron temperature in the presence of junction resistances $R_{pn}$ is larger than that in uniform channel by a factor of $1+6 R_{pn}/R_\mathrm{bulk}$. This fact can also contribute to the observed larger PR in the non-uniform channel, compared to the uniformly doped case.

The measured PC lets us to evaluate the performance of our device acting as a sub-THz bolometer. The power incident on the antenna  of 0.8 mm$^2$ area of device \#2 is estimated as $P\approx 40$~nW, which accounts for angular divergence of the beam and losses in lenses and intentionally introduced filters (Supplementary material, Sec. V). Under such conditions, the maximum photoresistance reaches $\delta R_\mathrm{THz}/R \approx 0.06$. This allows us to estimate the voltage responsivity $r_V$ and noise equivalent power
\begin{equation}
    {\rm NEP}^* = \frac{S_\mathrm V}{\delta R_{\rm THz} I_{\rm bias} / P},
\end{equation}
where $S_\mathrm V = S_{1/f} + S_\mathrm{JN}$ is the total voltage noise comprised of Flicker ($S_{1/f}$) and Johnson-Nyquist ($S_\mathrm{JN}$) components. At 34 Hz measurement frequency the {\it measured} noise $S_{\rm V} = 63$ nV/Hz$^{1/2}$, wich results in ${\rm NEP}^* = 47$~pW/Hz$^{1/2}$. NEP can be further reduced at higher modulation frequencies $f^* \gtrsim 15$~kHz, where the $1/f$-noise will surrender to thermal noise. This allows us to estimate the minimum achievable ${\rm NEP}_{\rm min}^* \sim 3$~pW/Hz$^{1/2}$. This is a good value for the bolometric class of photodetectors, it is comparable to low-temperature germanium bolometers \cite{low_Ge_1961}. The obtained values of $r_V$ and ${\rm NEP^*}$ at 25~K are of the same order as in \cite{yan_2012} for BLG under IR illumination in terms of external responsivity, but will become superior to reported previously if our device is cooled to the same temperature of 5~K.


In conclusion, our measurements have revealed a strong impact of electrically-induced $p$-$n$ junctions in bilayer graphene on its electrical conduction at cryogenic temperatures. Further combined studies of temperature-dependent resistivity and THz photoresitivity have revealed an unexpected increase in junction resistance with increasing $T$. We have shown that this anomaly can be explained with interband tunneling theory taking into account the thermal smearing of the Fermi surface. Instructively, the measured photoresistivity of BLG $p$-$n$ junctions is rather high, and leads to quite a large estimate of external responsivity $r_V \approx 1200$~V/W and low ${\rm NEP}^*$ of $3...47$~pW/Hz$^{1/2}$ at $T=25$~K.
 
\begin{suppinfo}
The Supporting Information contains following subsections: (I) Device fabrication, (II) Carrier concentration and bandgap in BLG, (III) Carrier mobility extraction, (IV) Device behaviour under inverted back gate, (V) NEP Calculation, (VI) Calculation of the junction tunneling resistance
\end{suppinfo}

\begin{acknowledgement}
The work of D.M., E.T., M.K., V.S. and D.S. (sample fabrication, electrical and THz measurements) was supported by the Russian Science Foundation, grant \# 21-79-20225. The devices were fabricated using the equipment of the Center of Shared Research Facilities (MIPT). S.Z. acknowledges the support of the  the Ministry of Science and Higher Education of the Russian Federation (grant \#  FSMG-2021-0005). E.T., M.K., I.S., and K.N. thank Vladimir Potanin via Brain and Conscionsness Research Center.
\end{acknowledgement}

\section*{Author declarations}

\subsection*{Conflict of interest}
The authors have no conflicts of interest to disclose.

\subsection*{Author contributions}
D.S. and D.B. conceived the idea of experiment; K.N., D.B., E.T. and M.K. established the technology of sample fabrication at MIPT;  M.K. fabricated the samples; D.M. and S.Z. designed the experiment; D.M., E.T., and V.S. performed experimental measurements; D.M. treated the experimental data; I.S. and D.S. developed the theoretical model; D.M., I.S. and D.S. wrote the text. All authors contributed to the discussions of data.

\subsection*{Data availability}
Original data are available from the corresponding author upon a reasonable request.

\bibliography{}

\end{document}


\renewcommand{\theequation} {S\arabic{equation}}
\renewcommand{\thefigure} {S\arabic{figure}}
	
\maketitle 
	

\section{I. Device fabrication}

Device was made by incapsulating of BLG between two slabs of hexagonal boron nitride (hBN) using dry transfer technique. This involved standard dry-peel technique to obtain graphene and relatively thick hBN crystals. The flakes were stacked on top of each other using a stamp made of PolyBisphenol carbonate (PC) on polydimethylsiloxane (PDMS) and deposited on top of an oxidized (280 nm of SiO$_2$) low-conductivity silicon wafer (KDB-12). The resulting  thickness of the top and bottom hBN was measured by atomic force microscopy. Then electron-beam lithography and reactive ion etching with SF6  (30 sccm, 125 Watt power) were employed to define contact regions in the obtained hBN/BLG/hBN heterostructure. Metal contacts to graphene were made by electron-beam evaporating 3 nm of Ti and 70 nm of Au. Afterwards, a second e-beam lithography was used to design the top split gate (3 nm of Ti, 70 nm of Au). The graphene channel was finally defined by a third round of e-beam lithography, followed by reactive ion etching using PMMA as the etching mask.

\section{II. Carrier concentration and bandgap in BLG}

The carrier density $n$ and the band gap $\Delta$ were calculated according to the theory from \cite{alymov_2016}. The interlayer permittivity of graphene was taken equal to 2, according to the latest experimental data \cite{icking_2022}. Then the following linear approximation agrees very well with this theory in the temperature range from 10 to 300~K:

\begin{equation}
n \approx \frac{\varepsilon_0}{e} \left(\frac{\varepsilon_T V_{TG}}{d_T} + \frac{\varepsilon_B V_{BG}}{d_B}\right)\;,
\label{eq:blg_n}
\end{equation}
\begin{equation}
\Delta \approx 0.05 \mathrm{\frac{eV}{V/nm}} \left|\frac{\varepsilon_T V_{TG}}{d_T} - \frac{\varepsilon_B V_{BG}}{d_B}\right|\;
\label{eq:blg_delta}
\end{equation}

where $\varepsilon_T$, $\varepsilon_B$ are relative permittivities of dielectrics between graphene and top and bottom gates, ${d_T}$, ${d_B}$ --- their thicknesses and $e$ --- electron charge. So, the carrier concentration is determined by the sum of the displacement fields, and the bandgap is determined by the their difference.

\section{III. Carrier mobility extraction}

\begin{figure}[h]
\includegraphics[width=0.6\textwidth]{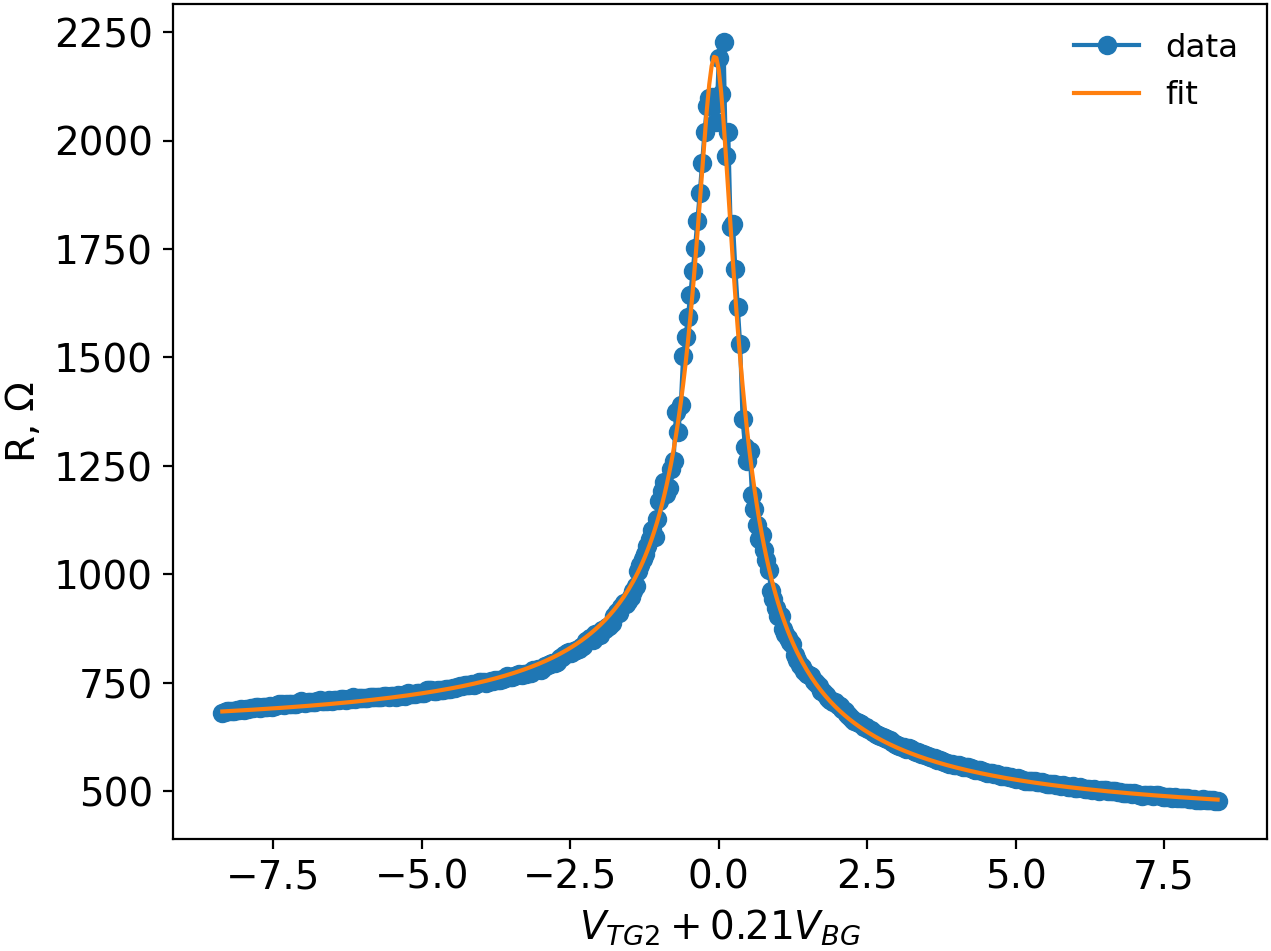}
\caption{Fitting the slice of map on Fig. 2(a) of main text along $\Delta=0$ by \cite{kim_fit_2009} that gives $\mu_e=29,000$~cm$^2$/V$\cdot$s and $\mu_h=33,000$~cm$^2$/V$\cdot$s.}
\label{fig:PN_junction}
\end{figure}

\section{IV. Device behaviour under inverted back gate}

\begin{figure}[h]
\includegraphics[width=0.8\textwidth]{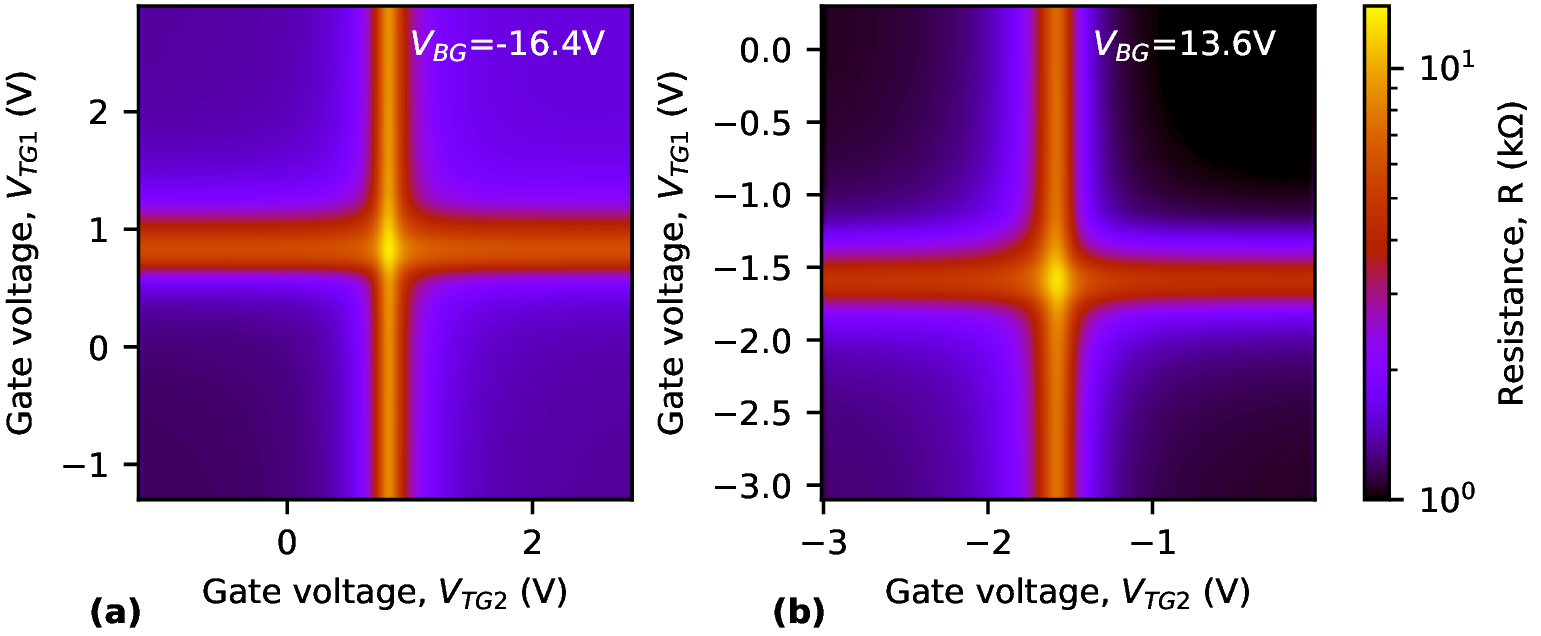}
\caption{Resistance map of device for two back gates of different signs.}
\label{fig:supp_map_R}
\end{figure}
\begin{figure}[h]
\includegraphics[width=0.8\textwidth]{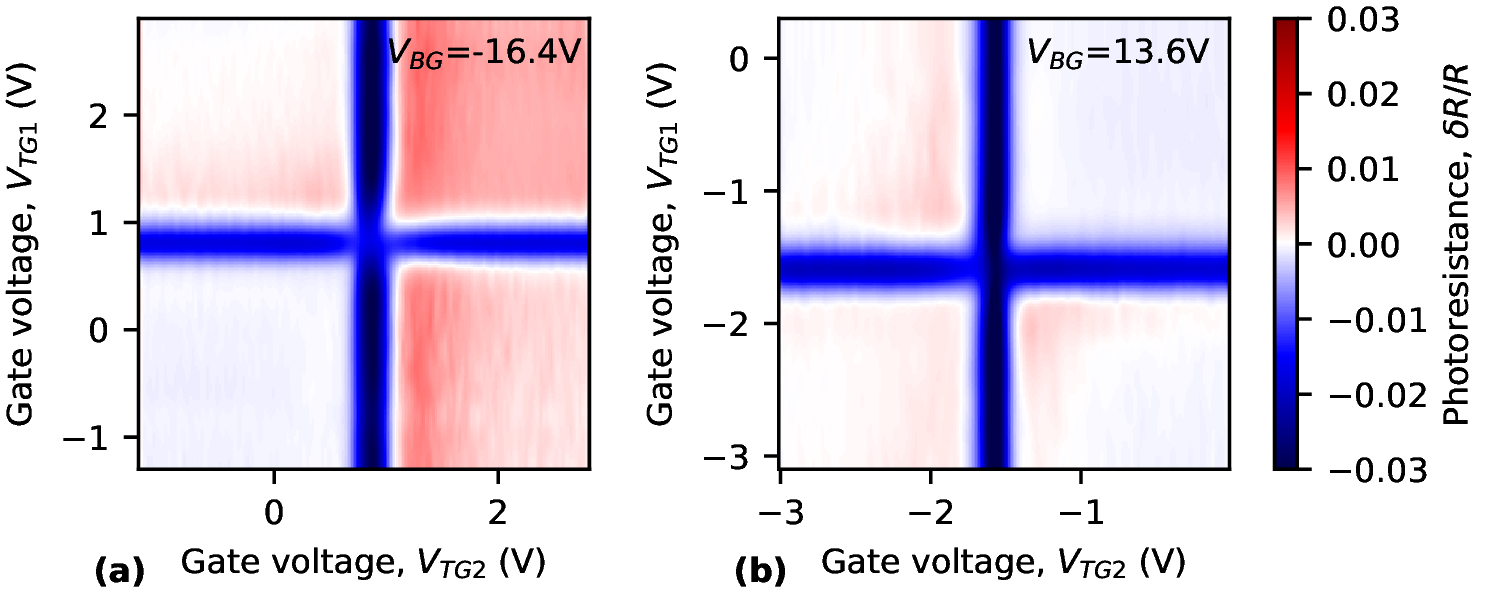}
\caption{Photoconductivity for two back gates of different signs.}
\label{fig:supp_map_dR}
\end{figure}
Figure \ref{fig:supp_map_R} shows resistance maps for device \#2.  The device differed from the device \#1 described in the main text only in the size of the graphene channel ($W=9\mu$m, $L=6\mu$m) and the thickness of hBN (30 and 50 nm for top and bottom slab). Figure \ref{fig:supp_map_dR} shows photoresistance maps for device \#2.

\section{V. NEP Calculation}

For the device described in the paper linear polarized light from a source with a power of 10 mW and divergence angle FWHM of 20$^o$, passes through a set of filters with a total transmittance of 0.003, polarizer mounted at 45$^o$ (transmittance 0.5), and fall on a silicon lens 12 mm in diameter. Lens is mounted on a sample substrate located in a cryostat at a distance of 17.5 ~cm from the source. We assume that the lens focuses all the incident light into a region smaller than or equal to the size of the device antenna (1 mm). The coefficient of reflection of light from the first surface of the Si lens is 26\%, and we neglect the reflection from the mylar film cryostat window. Total power that reaches the device is 0.5 $\mu$W.

We also have measured device \#2, which demonstrated higher NEP. In the experiment it was mounted at a distance of 19 cm from the same THz source. Here we used the same polarizer and a set of filters with total transmittance of 0.03. We didn't use silicon lens, and take the sample absorbtion cross section equal to antenna size (1 mm). It gives us illuminating power $P=42$~nW. Taking experimentally measured signal-to-noise ratio S/N = 700 at bandwidth of 1.6~Hz we have NEP* = 47~pW/Hz$^{1/2}$.

Because we did not take great measures to minimize noise and protect against interference, we will give an estimation of the achievable NEP. Firstly, measurements were done at frequency of 34 Hz. It is most likely that in our device the response is due to the heating of the electrons, and not the lattice, which is usually true for graphene-based devices \cite{yan_2012}. Therefore, the response will depend slightly on the measurement frequency up to MHz or GHz, but at these frequencies the $1/f$ noise will drop significantly and we will neglect it. The maximum of response is  near the CNP, where there are no $p$-$n$ junctions and the associated shot noise. The thermal fluctuation NEP* is $\sqrt{4kT^2/R_\mathrm{th}}=38$ fW/Hz$^{1/2}$ using $R_\mathrm{th}$ = $dT/dP=\frac{\delta R/R}{1/R\cdot dR/dT}\frac{1}{P}\approx 0.024$~K/nW. The Johnson–Nyquist noise is $\sqrt{4kTR}\approx$3.3~nV/Hz$^{1/2}$ and voltage responsivity $R_V=1200$~kV/W at current of 400~nA rms gives NEP*~= 3~pW/Hz$^{1/2}$ at $T=25$~K.

\section{VI. Calculation of the junction tunneling resistance }

Tunneling can be described by Leo Esaki's expression

\begin{equation}
    I\sim\int^{E_{v}}_{E_{c}}\rho_{c}(E)\cdot\rho_{v}(E)\cdot D(E)\cdot(f_{c}(E+eV)-f_{v}(E))dE,
\label{eq:Esaki_formula}
\end{equation}
cwhere $E_{c},E_{v}$ correspond to the energies of the bottom of the conduction band of the $n$-doped region and the top of the valence band of the $p$-doped region of the device, $f_{c},f_{v}$ are the Fermi-Dirac distributions in the $n$- and $p$-regions, respectively, $D(E)$ is the transparency of the barrier. The entire temperature dependence of the tunneling current is contained in the Fermi-Dirac distributions. Thus, for simplicity, one can set the transparency of the barrier to be independent of energy, thereby focusing on the temperature dependence. Since the Esaki formula gives only a proportional value of the current in the theoretical description, the proportionality constant was taken in such a way as to obtain the best approximation of the electric current from the experiment. The resistance of the $p$-$n$ junction was calculated as $R=V/I$.



\bibliography{}